\documentclass[superscriptaddress,showpacs,preprintnumbers,nofootinbib,twocolumn]{revtex4-1}

\usepackage{amsmath,amssymb}
\usepackage{graphicx}

\usepackage{color}

\newcommand{\CP}{$\mathcal{CP}$\,}

\newcommand{\Tr}{\mathrm{Tr}\,}

\begin{document}

\title{Invariant approach to \CP
in family symmetry models}
\author{Gustavo C. Branco}
\email{gbranco@tecnico.ulisboa.pt}
\affiliation{{\small Centro de F{\'\i}sica Te\'orica de Part{\'\i}culas - CFTP, \\
 Instituto Superior T\'ecnico - IST, Universidade de Lisboa - UL,\\
Avenida Rovisco Pais, 1049-001 Lisboa, Portugal}
}
\author{Ivo de Medeiros Varzielas}
\email{ivo.de@soton.ac.uk}
\affiliation{{\small Department of Physics, University of Basel,}\\
Klingelbergstr. 82, CH-4056 Basel, Switzerland}
\affiliation{{\small School of Physics and Astronomy, University of Southampton,}\\
Southampton, SO17 1BJ, U.K.}
\author{Steve F. King}
\email{s.f.king@soton.ac.uk}
\affiliation{{\small School of Physics and Astronomy, University of Southampton,}\\
Southampton, SO17 1BJ, U.K.}

\begin{abstract}
We propose the use of basis invariants, valid for any choice of \CP transformation,
as a powerful approach to studying specific models of 
\CP violation in the presence of discrete family symmetries. We illustrate the virtues of this approach for examples based on $A_4$ and $\Delta(27)$ family symmetries. For $A_4$, we show how to elegantly obtain several known results in the literature.
In $\Delta(27)$ we use the invariant approach to identify how {\em explicit} (rather than {\em spontaneous}) 
\CP violation arises, which is geometrical in nature, i.e. persisting for arbitrary couplings in the Lagrangian.
\end{abstract}
\maketitle

\CP symmetry, the combination of particle-antiparticle exchange and space inversion,
is known to be violated by the weak interactions involving quarks in the Standard Model (SM)
\cite{Christenson:1964fg}. 
The origin of the observed SM quark \CP violation can be traced to the existence of three generations of quarks with non-trivial weak mixing described by the complex CKM matrix \cite{Kobayashi:1973fv}. However, the CKM matrix can be parameterised in different ways, and it was later realised that the amount of \CP violation in physical processes always depends on a particular weak basis invariant which can be expressed in terms of the quark mass matrices
\cite{Jarlskog:1985ht,Bernabeu:1986fc}.

Although Sakharov taught us that \CP violation is a necessary condition for explaining the matter-antimatter asymmetry of the Universe \cite{Sakharov:1967dj}, it became clear that the
observed quark \CP violation
is insufficient for this purpose \cite{Kuzmin:1985mm}, motivating new sources of \CP violation beyond the SM. 
One example of such new physics is neutrino mass and mixing involving new \CP invariants \cite{Branco:1986gr}. 
Indeed, following the discovery of a sizeable leptonic reactor angle
\cite{An:2012eh}, it is possible that leptonic \CP violation could be observed in the foreseeable future through neutrino oscillations, making such questions 
particularly timely \cite{Fogli:2012ua}. 

In accommodating neutrino mass and lepton mixing, one is forced to extend the SM in some way.
A popular idea is that large leptonic mixing angles arise from
some discrete family symmetry (for a review see e.g. \cite{King:2013eh}). 
One possibility is to impose a specific CP symmetry which transforms generations non-trivially as in \cite{Ecker:1981wv} (see \cite{Grimus:2003yn} for more recent examples).
The interplay of discrete family symmetry and \CP symmetry leads to certain consistency relations 
which any theory must obey \cite{Feruglio:2012cw}.
Although the consistency relations have been widely used~\cite{Feruglio:2013hia,Ding:2013bpa}, the invariant approach \cite{Branco:1986gr} is often neglected.

The main purpose of this work is to illustrate, with a few examples, the utility and power of weak basis invariants 
 \cite{Branco:1986gr} in the analysis of concrete models of neutrino mass, mixing and \CP violation involving discrete family symmetry. 
We show that such an approach, which relies on a knowledge of the Lagrangian of the model, is complementary to the 
approach based on the consistency relations \cite{Feruglio:2012cw}. 
Indeed we will show how the consistency conditions
can be derived from the requirement that the Lagrangian is invariant under both \CP symmetry and the discrete family symmetry. 
Therefore, in analysing particular models, the use of 
weak basis invariants alone is both sufficient and convenient.

To illustrate the virtues of the invariant approach in analysing discrete family symmetry models of leptons,
it suffices to consider a couple of 
examples based on $A_4$ and $\Delta(27)$ family symmetries. For $A_4$, we show how to elegantly obtain several known results in the literature \cite{Ding:2013bpa} via the use of weak basis invariants.
In $\Delta(27)$ we use the invariant approach to identify how {\em explicit} geometrical \CP violation,
i.e. persisting for arbitrary couplings in the Lagrangian, arises. This is to be contrasted with {\em spontaneous} geometrical \CP violation \cite{Branco:1983tn,de:2011zw} where a \CP conserving Lagrangian undergoes 
\CP violation due to vacuum expectation values (VEVs). In both cases the term ``geometrical'' refers to the fact that the 
\CP violation is controlled by the complex phase $\omega \equiv e^{i 2 \pi/3}$ emerging from the order three generators of $\Delta(27)$.

It is worthwhile to first recap how the invariant approach works for any theory where the Lagrangian is specified.
Following \cite{Bernabeu:1986fc}, to study \CP symmetry in any model one divides
a given Lagrangian as follows,
\begin{equation}
\mathcal{L}=\mathcal{L_{CP}}+\mathcal{L}_{rem} \,,
\label{L}
\end{equation}
where $\mathcal{L_{CP}}$ is the part that automatically conserves \CP (like the kinetic terms and gauge interactions \footnote{Pure gauge interactions conserve \CP \cite{Grimus:1995zi}.}) while $\mathcal{L}_{rem}$ includes 
the \CP violating non-gauge interactions such as the Yukawa couplings.
Then one considers the most general \CP transformation that leaves $\mathcal{L_{CP}}$ invariant and check if invariance under \CP restricts  $\mathcal{L}_{rem}$ - only if this is the case can $\mathcal{L}$ violate \CP.

In the presence of a family symmetry $G$, one may check if 
a given vacuum leads to spontaneous \CP violation, as follows.
Consider a Lagrangian invariant under $G$ and \CP, containing a series of scalars which under \CP transform as 
$(\mathcal{CP}) \phi_i (\mathcal{CP})^{-1} =U_{ij}  \phi_j^*$. In order for the vacuum to be \CP invariant, the following relation has to be satisfied:
$ <0| \phi_i |0> = U_{ij} <0| \phi_j^* |0>$. The presence of $G$ usually allows for many choices for $U$. 
If (and only if) no choice of $U$ exists which satisfies the previous condition, will the vacuum violate \CP, leading to spontaneous \CP violation.
In order to prove that no choice of $U$ exists one can construct
\CP-odd invariants.

As a brief review of how to derive \CP-odd invariants, consider the Lagrangian of the leptonic part of the SM extended by Majorana neutrino masses. After electroweak breaking at low energies, the most general mass terms are:
\begin{equation}
\label{low}
 -\mathcal{L}_m=m_l  \overline{e}_L e_R +  \tfrac{1}{2} m_\nu \overline{\nu}_{L} \nu^{c}_L + H.c.\,,
\end{equation}
where $L= (e_L, \nu_L)$ stand for the left-handed neutrino and charged lepton fields in a weak basis and $e_R$ for the right-handed counterpart.
Due to the $SU(2)_L$ structure, the most general \CP transformation which leaves the leptonic gauge interactions invariant are:
\begin{equation}
(\mathcal{CP}) L (\mathcal{CP})^\dagger =iU \gamma^0 \mathcal{C} \bar{L}^T, \ \ 
(\mathcal{CP}) e_R (\mathcal{CP})^\dagger =iV \gamma^0 \mathcal{C} \bar{e}_R^T \,.
\label{LCP}
\end{equation}
In order for $\mathcal{L}_m$ to be \CP invariant, under Eq.(\ref{LCP}) the terms shown in the Eq.(\ref{low}) go into the respective $H.c.$ and vice-versa:
\begin{equation}
U^\dagger  m_{\nu} U^* = m_{\nu}^*, \ \ \ \ 
U^\dagger m_{l} V = m_{l}^* \,.
\label{mlCP}
\end{equation}
From Eq.(\ref{mlCP}) one can infer how to build combinations of the mass matrices that will result in equations where $U$ and $V$ cancel entirely.
For any number of generations we have \cite{Bernabeu:1986fc}:
\begin{equation}
I_1 \equiv \Tr \left[H_\nu , H_l \right]^3 = 0\,,
\label{hhcube}
\end{equation}
where $H_\nu \equiv m_\nu m_\nu^\dagger$ and $H_l \equiv m_l m_l^\dagger$.
This equation is a necessary condition for \CP invariance,
encoding having no Dirac-type \CP violation. It can also be shown to be a sufficient condition in the case of 3 generations,
which we will do when discussing $A_4$ later.
The low-energy limit of the leptonic sector with 3 Majorana neutrinos has also two Majorana-type \CP violating phases, and it turns out
there are 3 necessary and sufficient conditions for low energy leptonic \CP invariance: in addition to Eq.(\ref{hhcube}), two more  \CP-odd invariants can be defined \cite{Branco:1986gr}, which we shall not consider further here.

In this work we are interested in applying these ideas to models of leptons involving discrete family symmetry.
The first point we wish to make is that, once a Lagrangian is specified, which is invariant under a family symmetry $G$
and some \CP transformation, then the consistency relations \cite{Feruglio:2012cw} are automatically satisfied.
In order to prove this it is sufficient to 
consider some generic 
Lagrangian invariant under a family symmetry transformation, involving some mass term $m$
(Dirac or Majorana), then define
$H=m m^{\dagger}$.
Under some $G$ transformation, 
$\rho(g)$, the mass term remains unchanged implying:
\begin{equation}
\rho(g)^{\dagger}H\rho(g)=H.
\label{cond00}
\end{equation}
Invariance of the Lagrangian under \CP transformation $U$ requires the mass term to swap with its $H.c.$, hence:
\begin{equation}
U^{\dagger}H U=H^*
\label{cond0}
\end{equation}
Taking the complex conjugate of Eq.(\ref{cond00}) we find,
\begin{equation}
(\rho(g)^{\dagger})^*H^*\rho(g)^*=H^*=U^{\dagger}HU,
\label{cond01}
\end{equation}
using Eq.(\ref{cond0}) for the last equality. 
Using Eq.(\ref{cond0}) again:
\begin{equation}
(\rho(g)^{\dagger})^*U^{\dagger}HU\rho(g)^*=U^{\dagger}HU.
\label{cond02}
\end{equation}
Hence by using once more Eq.(\ref{cond00}) for a $g'$, we finish with:
\begin{equation}
U(\rho(g)^{\dagger})^*U^{\dagger}HU\rho(g)^*U^{\dagger}=H=\rho(g')^{\dagger}H\rho(g').
\label{cond03}
\end{equation}
By comparing both sides of Eq.(\ref{cond03}) we identify:
\begin{equation}
U\rho(g)^*U^{\dagger}=\rho(g')
\label{consistency}
\end{equation}
which is just the consistency relation  \cite{Feruglio:2012cw}.
In other words, if we consider Eqs.(\ref{cond00}) and (\ref{cond0}) we do not need to consider the consistency
condition separately since it always follows.

We now move onto our first illustrative example, based on $G=A_4$
(see e.g. \cite{King:2011zj} for the basis choice and conventions).
To proceed with the invariant approach we consider the $A_4$ invariant Yukawa Lagrangian of a leptonic sector containing fields in all possible representations of $A_4$:
lepton doublets
$L=(\nu_{lL}, l_L)=3$, where $l=e,\mu,\tau$,
charged leptons
$e^c=1$, $\mu^c=1''$, $\tau^c=1'$,
Higgs flavons
$\varphi_{S} =3$, $\varphi_T=3$,
$\xi=1$, $\xi'=1'$, $\xi''=1''$.
\begin{eqnarray}
&{\cal L}_{A_4}=
-y_e(L \varphi_T )_1 \,e^c -  y_{\mu}(L \varphi_T)_{1'} \,\mu^c -y_{\tau}(L \varphi_T)_{1''} \,\tau^c  
\nonumber \\
&-\frac{y_1}{2}\varphi_{S}(LL)_{3_s} - \frac{y_2}{2}\xi (LL)_1-\frac{y_3'}{2}\xi' (LL)_{1''}
-\frac{y_3''}{2}\xi'' (LL)_{1'} \nonumber \\
&+H.c. 
\label{LA4}
\end{eqnarray}
Here $(\cdots)_{r}$ denotes the $A_4$ contraction into
representation ${\bf{r}}$. 
The only Higgs which can get a VEV without breaking $A_4$ is
$\langle \xi \rangle$.
Giving it a VEV leads to a very simple neutrino mass matrix in unbroken $A_4$,
from the $(LL)_1$ contraction:
\begin{equation}
m_{\nu}^0=\beta  \begin{pmatrix}1&0&0\\0&0&1\\0&1&0 \end{pmatrix} \,, \quad \beta= (y_2  \langle  \xi \rangle)^*\,.
\label{mnu0}
\end{equation}
Defining
$H^0_{\nu}=m^0_{\nu}m_{\nu}^{0 \dagger} = |\beta|^2I$,
we get that $H^0_\nu$ is trivially invariant under \CP: 
\begin{equation}
U^{\dagger}H^0_{\nu}U=H_{\nu}^{0*},
\label{cond001}
\end{equation}
for any unitary matrix $U$.
For $m_{\nu}^0$:
\begin{equation}
U^{\dagger}m_{\nu}^0U^*=m_{\nu}^{0*},
\label{cond002}
\end{equation}
\CP conservation can be seen by using $U=e^{i \arg(\beta)}\rho_{\mathbf{3}}(g)$.
Having complex $\beta$ is consistent with \CP invariance, and the existence of one \CP transformation proves the Lagrangian
respects \CP. 
The invariant approach for the single allowed mass term can only lead to  \CP-odd invariants of the form
\begin{equation}
\Im \Tr [ (m_{\nu}^{0 \dagger}m_{\nu}^0)^{n_1*} (m_{\nu}^0m_{\nu}^{0 \dagger})^{n_2} (m_{\nu}^{0 \dagger}m_{\nu}^0)^{n_3*} (...)]
\end{equation}
where $n_i$ are positive integers.
All these  \CP-odd invariants vanish because of Eq.(\ref{mnu0}), so we conclude without much effort that \CP invariance is inevitable and \CP is automatically conserved for this Lagrangian with unbroken $A_4$.

What about the \CP transformation of the other terms in the Lagrangian? It is possible to consistently define
a \CP transformation for all terms in Eq.(\ref{LA4}),
e.g. $U=I$ for the triplets, with a suitable and different phase for each triplet field,
and a phase for each singlet field. The phases are chosen with respect to the phases of the couplings which can all be complex.
This is both because in Eq.(\ref{LA4}) a single matrix structure for $U$ works for all the Yukawa structures involving the triplets, and because there is a different field
for each coupling.
Therefore it is not true that all $A_4$ invariant Lagrangians are \CP invariant:
adding the term $a\xi'\xi''$ to Eq.(\ref{LA4}) leads to \CP violation for complex $a$.
This illustrates that \CP need not be conserved for $A_4$ invariant Lagrangians.

When $\varphi_{S},\varphi_T,\xi,\xi',\xi''$
acquire VEVs, $A_4$ is broken.
We consider now realistic models with different subgroups preserved in the neutrino and charged lepton sectors and investigate the conditions for \CP conservation. We assume the VEVs \cite{King:2011zj},
\begin{equation}
\langle \varphi_S \rangle = {v_S}
\begin{pmatrix}1\\ 1\\1 \end{pmatrix} \ , \qquad
\langle \varphi_T \rangle =v_T \begin{pmatrix}1\\ 0\\0 \end{pmatrix}  \ , 
\label{eq-AFalignment}
\end{equation}
where $S\langle \varphi_S \rangle = \langle \varphi_S \rangle$ hence $\langle \varphi_S \rangle$ leaves $S$ unbroken,
while $T\langle \varphi_T \rangle = \langle \varphi_T \rangle$ hence $\langle \varphi_T \rangle$ leaves $T$ unbroken.
In the neutrino sector $S$ is preserved, the previous matrix $m_{\nu}^0$ becomes enlarged to:
\begin{equation}
m_{\nu}=m_{\nu}^0+ \alpha \begin{pmatrix} 2&-1&-1\\-1&2&-1\\-1&-1&2\end{pmatrix}
+\gamma \begin{pmatrix}0&0&1\\0&1&0\\1&0&0 \end{pmatrix}
+\delta \begin{pmatrix} 0&1&0\\1&0&0\\0&0&1\end{pmatrix},
\label{mR}
\end{equation} 
where $\alpha= (y_1v_S)^*$,
$\gamma=(y_3' \langle  \xi' \rangle)^*$,  $\delta=(y_3'' \langle  \xi'' \rangle)^*$.
The charged lepton mass matrix $m_l$ preserves $T$ and is diagonal,
$m_{l}={\rm diag}(m_e,m_{\mu},m_{\tau})$
where $m_e=(y_ev_T)^*$, $m_{\mu}=(y_{\mu}v_T)^*$, $m_{\tau}=(y_{\tau}v_T)^*$.

With $H_l$ diagonal, $I_1$ is
\begin{equation}
I_1
=6i(m_{\mu}^2-m_e^2)(m_{\tau}^2-m_e^2)(m_{\tau}^2-m_{\mu}^2)\Im (H^{21}_{\nu}H^{13}_{\nu}H^{32}_{\nu}).
\end{equation}
\CP conservation forces $I_1=0$ and since there are no mass degeneracies,
with the off-diagonal phases summing to zero (modulo integer multiples of $\pi$),
$\phi_{21}+\phi_{13}+\phi_{32}=0$
(where we denoted the phases of $H^{ij}_{\nu}$ as $\phi_{ij}$),
we find
\begin{equation}
\Im (H^{21}_{\nu}H^{13}_{\nu}H^{32}_{\nu})=-\Im (\beta \delta^* + \gamma \beta^* + \delta \gamma^*) \Re (R)
\label{I1A4}
\end{equation}
where $R$ is a rather complicated expression,
\begin{eqnarray}
R&=&27|\alpha|^4-6|\alpha|^2|\beta + \gamma + \delta|^2+|\gamma \delta|^2+|\delta \beta|^2+|\beta \gamma |^2
\nonumber \\
&+& 4|\beta|^2 (\gamma \delta^*)+4|\gamma|^2 (\delta \beta^*)+4|\delta|^2 (\beta \gamma^*)\nonumber \\
&+&-6\alpha^{*2}(\beta^2+\gamma^2+\delta^2-\beta \gamma - \delta \beta - \gamma \delta)\nonumber \\
&+& 2\beta^{*2}(\gamma^2+\delta^2+\gamma \delta) 
+ 2\gamma^{*2}(\delta^2+\delta \beta) + 2\delta^{*2}\beta \gamma  \nonumber .
\end{eqnarray}
From Eq.(\ref{I1A4}) we learn that setting to zero any two of the
parameters $\beta, \gamma, \delta$ (dropping any two of the singlets)
automatically leads to $I_1=0$ for any values of the remaining parameters and leading to the absence of Dirac-type \CP violation. Indeed this coincides with what is known in the literature, since at least one non-trivial singlet is required to obtain non-vanishing reactor angle with this Lagrangian.

For the \CP conserving cases,
the condition for \CP conservation is
\begin{equation}
U^{\dagger}H_{\nu}U=H_{\nu}^*
\label{cond}
\end{equation}
We find that, since $H_{\nu}$ is a Hermitian matrix whose off-diagonal phases sum to zero,
one solution to eq.\ref{cond} is
\begin{equation}
U'= {\rm diag}(e^{2i\phi_1},e^{2i\phi_2},e^{2i\phi_3})
\end{equation}
where the off-diagonal phases of $H_{\nu}$ are given by $\phi_{ij}=\phi_i-\phi_j$.
In fact,
it is always possible to remove the off-diagonal phases in $H_{\nu}$ completely by using charged lepton phase rotations
$L\rightarrow {\rm diag}(e^{-i\phi_1},e^{-i\phi_2},e^{-i\phi_3})L$
where the off-diagonal phases of $H_{\nu}$ are given by $\phi_{ij}=\phi_i-\phi_j$
as before.
In this basis, the \CP conserving $H_{\nu}$ is real and the \CP transformation 
in Eq.(\ref{cond}) is the unit matrix $U^I=I$.
Since $S$ is a conserved symmetry of the neutrino mass matrix,
$SH_{\nu}S=H_{\nu}$, it follows that also the following \CP transformation must also be possible, ${U'}^{S}= S U'$
or in the basis where $H_{\nu}$ is real, simply
${U}^{S}= S$.
It is interesting to compare the invariant approach (above) to that previously followed for the same
$A_4$ model \cite{Ding:2013bpa}, where the same results were obtained from the consistency condition.

We will now use the invariant approach to show for the first time how one obtains explicit geometrical \CP violation - i.e. \emph{\CP is explicitly violated by a phase only originating from the group structure, and not from arbitrary couplings}. We consider $G=\Delta(27)$, which can produce complex VEVs that lead to spontaneous geometrical \CP violation \cite{Branco:1983tn, de:2011zw}.
There are 12 \CP transformations consistent with $\Delta(27)$ triplets \cite{Nishi:2013jqa}, but to use the invariant approach it is sufficient to know how to build $\Delta(27)$ invariants.

$\Delta(27)$ has 3 generators but we need use only two of them here: $c$ (for cyclic) and $d$ (for diagonal), $c^3=d^3=\mathrm{I}$. It has 9 singlets which we label as $1_{ij}$ with $c$,$d$ represented by $c_{1_{ij}}=\omega^i$ and $d_{1_{ij}}=\omega^j$ ($\omega \equiv e^{i 2 \pi/3}$).
There are two $\Delta(27)$ triplets which we take as $3_{01}$ and $3_{02}$. $c$ is represented equally for both, but not $d$:
\begin{equation}
c_{3_{ij}}=
\begin{pmatrix}
	0 & 1 & 0 \\
	0 & 0 & 1 \\
	1 & 0 & 0
\end{pmatrix} ; d_{3_{ij}}=
\begin{pmatrix}
	\omega^i & 0 & 0 \\
	0 & \omega^j & 0 \\
	0 & 0 & \omega^{-i-j}
\end{pmatrix}
\end{equation}
In $\Delta(27)$, $3_{01} \otimes 3_{02} = \sum_{i,j} 1_{ij}$, and with
$A=(a_1,a_2,a_3)_{01}$ transforming as triplet $3_{01}$ and $\bar{B}=(\bar{b}_1,\bar{b}_2,\bar{b}_3)_{02}$ transforming as (anti-)triplet $3_{02}$,
the explicit construction of the singlets we require are
\begin{align}
(A \bar{B})_{00}&=(a_1 \bar{b}_1 + a_2 \bar{b}_2 + a_3 \bar{b}_3)_{00} \label{AB00} \\
(A \bar{B})_{01}&=(a_2 \bar{b}_1 + a_3 \bar{b}_2 + a_1 \bar{b}_3)_{01} \label{AB01} \\
(A \bar{B})_{02}&=(a_1 \bar{b}_2 + a_2 \bar{b}_3 + a_3 \bar{b}_1)_{02} \label{AB02} \\
(A \bar{B})_{20}&=(a_1 \bar{b}_1 + \omega a_2 \bar{b}_2 + \omega^2 a_3 \bar{b}_3)_{20} \label{AB20}
\end{align}
This can be verified by acting on the triplets with the generators.
The study of \CP in the context of $\Delta(27)$ with more singlets is a rich topic where the invariant approach proves to be extremely useful and we will present a more detailed exploration of it in a subsequent publication.

$\Delta(27)$ was first used for the lepton sector in \cite{deMedeirosVarzielas:2006fc}. We introduce now the SM fermions $L \sim 3_{01}$ and also $\nu^c \sim 3_{02}$.
In order to make this model physical, we complete it with a charged lepton Lagrangian that gives them diagonal mass matrix with a VEV that breaks $\Delta(27)$ for $\phi \sim 3_{02}$, $\langle \phi \rangle \propto (1,0,0)$:
\begin{equation}
-y_e (L \phi)_{00} \,e^c_{00} -  y_{\mu}(L \phi)_{01} \,\mu^c_{02} -y_{\tau}(L \phi)_{02} \,\tau^c_{01}  +H.c.
\label{D27cl}
\end{equation}
By using the invariant approach we found an interesting case for 3 $h_{ij}$ scalars in the neutrino sector, e.g.:
\begin{equation*}
\mathcal{L}_{3s} = y_{00} (L \nu^c)_{00} h_{00} + y_{01} (L \nu^c)_{02} h_{01} + y_{10} (L \nu^c)_{20} h_{10} + H.c.
\end{equation*}
In this Lagrangian $\Delta(27)$ remains unbroken until the $h_{ij}$ acquire VEVs.
The most general CP transformations are associated respectively to unitary transformations:
\begin{align*}
&h_{00} \rightarrow e^{i p_{00}} h_{00}^* ;\quad
h_{01} \rightarrow e^{i p_{01}} h_{01}^*; \quad
h_{10} \rightarrow e^{i p_{10}} h_{10}^*; \quad
\\ 
&L \rightarrow U_L^T L^* ;\quad
\nu^c \rightarrow U_\nu \nu^{c *}\,,
\end{align*}
such that, if we assume \CP invariance we have for the Yukawa matrices $Y_{ij}$ associated with each term
\begin{align}
U_L Y_{ij} U_\nu e^{i p_{ij}}= Y_{ij}^* \,,
\label{Y3s}
\end{align}
where $\Delta(27)$ invariance imposes $Y_{00}=y_{00} \mathrm{I}$ and
\begin{equation}
Y_{01}=y_{01}
\begin{pmatrix}
0 & 1 & 0\\
0 & 0 & 1\\
1 & 0 & 0
\end{pmatrix}  ; \quad
Y_{10}=y_{10}
\begin{pmatrix}
1 & 0 & 0\\
0 & \omega & 0\\
0 & 0 & \omega^2
\end{pmatrix} \,.
\end{equation}
If we solve Eq.(\ref{Y3s}) with $y_{ij} \neq 0$ we find no solution for either $U_L$ or $U_\nu$. We conclude that in this Lagrangian with unbroken $\Delta(27)$ \CP is violated in general and build:
\begin{equation}
I_{3s} \equiv \Im \Tr (Y_{00} Y_{01}^\dagger Y_{10} Y_{00}^\dagger Y_{01} Y_{10}^\dagger) \,.
\label{I3s}
\end{equation}
This \CP-odd invariant is sensitive to the presence of 3 scalars and
\begin{equation}
I_{3s}=\Im(3 \omega^2 |y_{00}|^2  |y_{01}|^2 |y_{10}|^2)
\end{equation}
where the only phase present is $\omega^2$. The invariant approach therefore shows that we have for the first time found a case where \emph{\CP is explicitly violated by a phase only originating from the group structure, and not from arbitrary couplings}.
This falls under the definition of geometrical \CP violation, but to distinguish it from already known cases where it occurs spontaneously, we refer to this as explicit geometrical \CP violation.

The mass structure for the Dirac neutrinos, when $\Delta(27)$ is broken and $a_{ij} = y_{ij} \langle h_{ij} \rangle$, is:
\begin{equation}
m_\nu=
\begin{pmatrix}
a_{00}+a_{10} & a_{01} & 0\\
0 & a_{00} + \omega a_{10} & a_{01}\\
a_{01} & 0 & a_{00} + \omega^2 a_{10}
\end{pmatrix} \,.
\end{equation}
We have 6 parameters ($a_{ij}$ being 3 complex numbers) and fix them to give 3 different neutrino masses and mixing angles (the charged leptons are diagonal). We have a prediction for the $\delta$ CP violating phase of the leptons, which we express in terms of
$I_1 \neq 0$ because:
\begin{equation}
\Im (H_{\nu}^{21} H_{\nu}^{13} H_{\nu}^{32})= \Im (a_{00}^3+a_{10}^3) (a_{01}^*)^{3} \,.
\end{equation}
This source of \CP violation depends on the relative phases of the parameters, but is predicted once the parameters are fixed to give the correct masses and mixings.

Is there a physical process where the explicit geometrical \CP violation could be probed? In principle yes. For this model, strictly from counting the number of Yukawa in $I_{3s}$, it could be probed in decays of the scalars $h_{ij}$ due to the interference of tree level and 2-loop processes. Because smaller  \CP-odd invariants involving $h_{ij}$ are automatically zero, lower order contributions are \CP conserving.

To summarise, the invariant approach is a powerful tool in the study of the \CP properties of specific Lagrangians, whether they are invariant under a family symmetry or not. We have demonstrated how it elegantly gives the relevant results for an $A_4$ framework. Then, in a realistic model of leptons with $\Delta(27)$, we obtained the strength of Dirac-type \CP violation and identified a case with explicit geometrical \CP violation.

\acknowledgments{
This project has received funding from the Swiss National Science Foundation. 
This project has received funding from the European Union's Seventh Framework Programme for research, technological development and demonstration under grant agreement no PIEF-GA-2012-327195 SIFT.
The authors also acknowledge partial support from the European Union FP7 ITN-INVISIBLES (Marie Curie Actions, PITN- GA-2011- 289442),
and by Funda\c{c}\~{a}o para a Ci\^{e}ncia e a Tecnologia
(FCT, Portugal) through the projects CERN/FP/123580/2011, PTDC/FIS-NUC/0548/2012, EXPL/FISNUC/
0460/2013, and CFTP-FCT Unit 777 (PEst-OE/FIS/UI0777/2013) which are partially funded through
POCTI (FEDER), COMPETE, QREN and EU.
We thank CERN for hospitality.
}

\end{document}